\documentstyle[aaspp4,11pt]{article}
\begin{document}

\title{Global Star Formation Rates in Disk Galaxies and Circumnuclear
Starbursts from Cloud Collisions} \author{Jonathan C. Tan}
\affil{Department of Astronomy, University of California, Berkeley, CA
94720\\email: jt@astron.berkeley.edu}

\section*{Abstract}

We invoke star formation triggered by cloud-cloud collisions to
explain global star formation rates of disk galaxies and circumnuclear
starbursts. Previous theories based on the growth rate of
gravitational perturbations ignore the dynamically important presence
of magnetic fields. Theories based on triggering by spiral density
waves fail to explain star formation in systems without such
waves. Furthermore, observations suggest gas and stellar disk
instabilities are decoupled. Following Gammie, Ostriker \& Jog (1991),
the cloud collision rate is set by the shear velocity of encounters
with initial impact parameters of a few tidal radii, due to
differential rotation in the disk. This, together with the effective
confinement of cloud orbits to a two dimensional plane, enhances the
collision rate above that for particles in a three dimensional box. We
predict $\Sigma_{SFR}(R)\propto \Sigma_{gas} \Omega (1-0.7
\beta)$. For constant circular velocity ($\beta = 0$), this is in
agreement with recent observations (Kennicutt 1998). We predict a
B-band Tully-Fisher relation: $L_{B}\propto v_{circ}^{7/3}$, also
consistent with observations. As additional tests, we predict enhanced
star formation in regions with relatively high shear rates, and lower
star formation efficiencies in clouds of higher mass.

Subject headings: galaxies: starburst and spiral --- ISM: clouds --- stars: formation 

\section{Introduction}

Understanding how the global star formation rates (SFRs) of galaxies
and starbursts depend on their physical properties is essential for an
understanding of galaxy evolution. Furthermore, such knowledge can
also reveal much about the star formation process itself.

Empirically, in disk galaxies (Kennicutt 1989 \& 1998, hereafter K89
\& K98) and the circumnuclear disks of starbursts (Kenney 1997; Downes
\& Solomon 1998, hereafter DS98), star formation occurs in regions
where the gas disk is unstable to gravitational perturbation
growth. This can be expressed as a condition on the surface density of
gas:
\begin{equation}
\label{Q}
\Sigma_{gas}>\Sigma_{crit}=\frac{\alpha \kappa \sigma_{gas}}{\pi G}\equiv Q \Sigma_{gas},
\end{equation}
(Toomre 1964, Quirk 1972), where $\sigma_{gas}$ is the gas velocity
dispersion; $\alpha$ is a dimensionless constant near unity, to
account for deviations of real disks from the idealized Toomre thin
disk, single fluid model; $Q$ is a dimensionless parameter and
$\kappa$ is the epicyclic frequency:
\begin{equation}
\label{kappa}
\kappa=\sqrt{2}\frac{v_{circ}}{R}\left(1+\frac{R}{v_{circ}}\frac{dv_{circ}}{dR}\right)^{1/2}=\sqrt{2}\frac{v_{circ}}{R}(1+\beta)^{1/2}.
\end{equation}
$v_{circ}$ is the circular velocity at a particular galactocentric
radius $R$, and $\beta\equiv d\:{\rm ln}v_{circ}/d\:{\rm ln}R$, which
is $0$ for a flat rotation curve. From the outermost galactic star
forming regions, K89 finds $\alpha\simeq0.67$, assuming
$\sigma_{gas}=6\:{\rm km/s}$. $\alpha<1$ is expected, because of the
destabilizing influence of a stellar disk (Jog \& Solomon 1984, Jog
1996). Where $Q<1$, for a certain range of scales, the gas disk is
gravitationally unstable, and fragments into bound clouds. When stars
form, the energy they release raises $\sigma_{gas}$ and star formation
is hypothesized (e.g. Silk 1997) and observed (K89, DS98) to
self-regulate, so that $Q\sim {\cal O}(1)$.

All star formation is observed to occur in molecular clouds, and the
majority in giant molecular clouds (GMCs), with masses $\gtrsim
10^{5}\:{\rm M_{\odot}}$ (see Blitz \& Williams 1999 and McKee 1999
for reviews). However, K89 reported the surprising result that the
correlation of SFR with the surface density of molecular gas was much
weaker than with the total (atomic + molecular). Uncertainties in CO
to ${\rm H_2}$ conversion may account for some of the poor
correlation, however, the data suggest the immediate supply of gas
controlling the SFR is both atomic and molecular. This implies the
atomic to molecular conversion timescale, $t_{conv}$, is short
compared to the timescale on which star formation is
regulated. Spitzer (1978) finds the rate constant for molecule
formation on dust grains to be approximately $2.0\times 10^{-17}\:{\rm
cm^{3}\:s^{-1}}$, for typical Galactic interstellar medium (ISM)
metallicities. Ignoring destruction processes, a naive estimate of the
time to convert a region with $n_{HI}\sim 1000\:{\rm cm^{-3}}$, perhaps
created from the collision of two atomic clouds, to H$_2$, gives
$t_{conv}\sim 2\times 10^{6}\:{\rm yrs}$, which is a relatively short
timescale.

\begin{figure}
\plotone{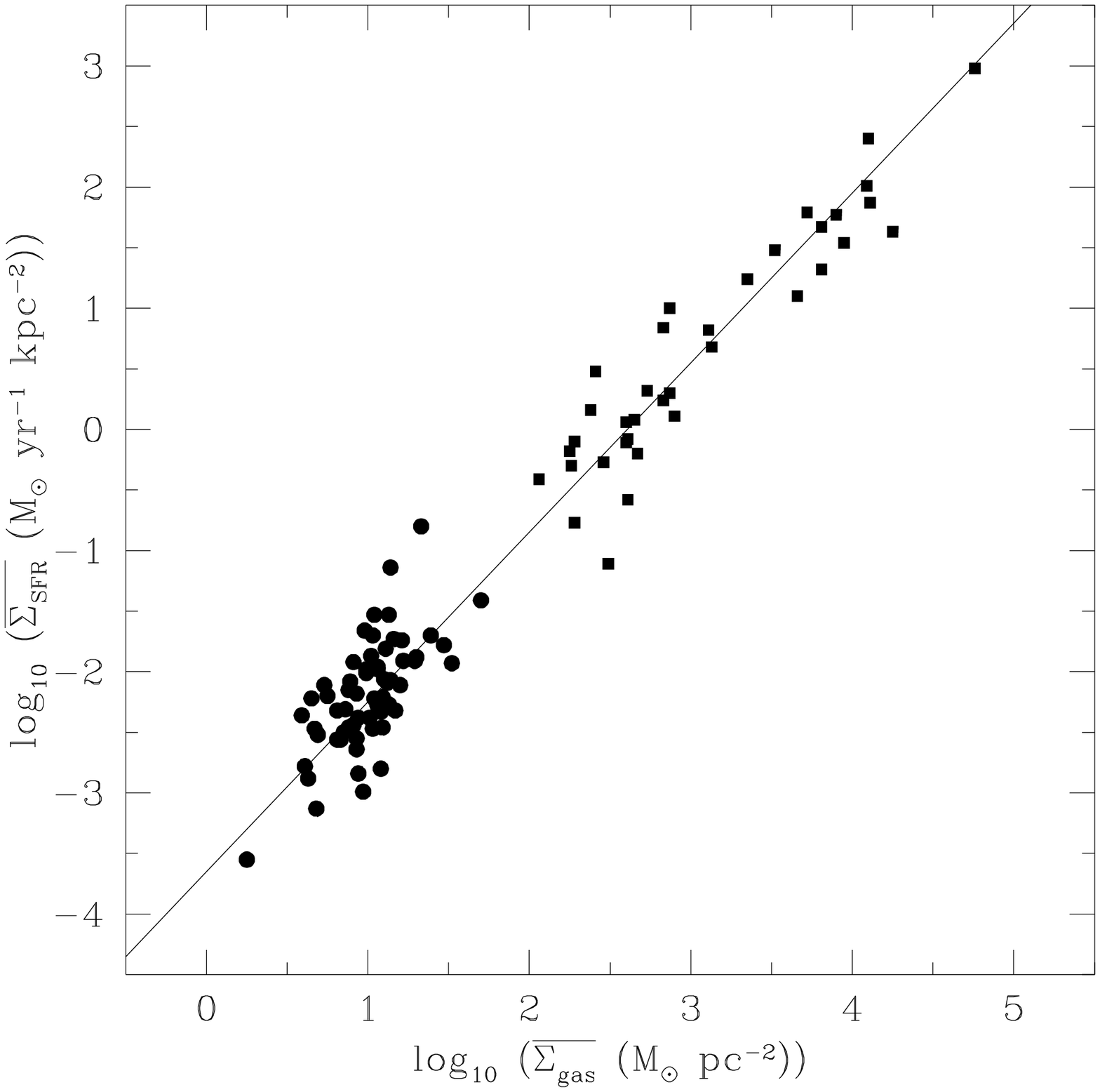}
\figcaption{
Classical Schmidt law : $\overline{\Sigma_{SFR}}\propto (\overline{\Sigma_{gas}})^{N}$. From Kennicutt (1998). Data are disk averaged quantities for normal galactic disks ({\it solid circles}) and circumnuclear starburst disks ({\it solid squares}). The line is a least-squares fit with index $N=1.40$. Systematic uncertainties between the normalization of the normal and starburst samples are of the order of a factor of two.
\label{fig:ken1}}
\end{figure}

Where $Q\lesssim 1$, the SFR is observed to be correlated with gas
density. Schmidt (1959) introduced the parameterization of the volume
densities $\rho_{SFR}\propto \rho_{gas}^n$, with $n\sim 1-2$. By
looking at about one hundred different galactic and circumnuclear
starburst disk systems, K98 found a similar relation for disk averaged
surface densities of gas and star formation, valid over five orders of
magnitude in $\overline{\Sigma_{gas}}$,
\begin{equation}
\label{kenlaw}
\overline{\Sigma_{SFR}}\propto\overline{\Sigma_{gas}}^{N},
\end{equation}
with $N\sim1.4\pm0.15$ (figure \ref{fig:ken1}) (however, see Taniguchi
\& Ohyama 1998). K98 finds the SFR is also correlated with the orbital
angular frequency, $\Omega$, via
\begin{equation}
\label{larson}
\overline{\Sigma_{SFR}}\propto\overline{\Sigma_{gas}}\Omega,
\end{equation}
(figure \ref{fig:ken2}). $\Omega$ is measured at the outer radius of
the star forming region (see \S\ref{diskav}).

Previous theories for explaining these relations fall into two broad
categories, based either on the growth rate of gravitational
perturbations in a disk or on the triggering of star formation in gas
passing through spiral or bar density waves. In this paper we present
a third paradigm, in which cloud collisions determine the SFR.

The timescale for perturbation growth can be expressed as
$\tau_{grow}\propto (G \rho)^{-0.5}$ (e.g. Larson 1988, 1992;
Elmegreen 1994; Wang \& Silk 1994), and so $\rho_{SFR}
\propto\rho_{gas}/\tau_{grow}\propto\rho_{gas}^{1.5}$. Assuming a
constant disk scaleheight, we obtain equation (\ref{kenlaw}) with
$N=1.5$ for local surface densities. However, disk averaged quantities
will depend on the radial gas distribution. We can also express
$\tau_{grow} \sim \alpha \sigma_{gas}/(\pi G \Sigma_{gas}) \sim
Q/\kappa$. Perturbation growth via swing amplification in a
differentially rotating disk grows in a similar manner (e.g. Larson
1988). By assuming star formation self-regulates and keeps $Q$
constant, Larson (1988) and Wang \& Silk (1994) predict
$\Sigma_{SFR}\propto \Sigma_{gas}/\tau_{grow} \propto\Sigma_{gas}
\Omega$, since $\kappa \propto \Omega$, for disks with flat rotation
curves.

However, these theories neglect the effects of magnetic fields and the
viscosity of the ISM. Gammie (1996) finds these significantly reduce
the growth rate of non-axisymmetric perturbations for typical Galactic
conditions. Furthermore, most Galactic disk stars form in localized,
highly clustered regions (Lada, Strom \& Myers 1993) in GMCs, and most
of the gas in the disk, including most of the bound gas, is not
directly involved in the star formation. In GMCs static magnetic
fields play a dynamically important role (e.g. McKee 1999, Heiles et
al 1993). Their presence sets a critical mass,
\begin{equation}
\label{M_B}
M_B=512\frac{\overline{B}^{3}_{1.5}}{\overline{n}^{2}_{H3}}\:{\rm M_{\odot}},
\end{equation}
(Bertoldi \& McKee 1992) for spherical clouds, where
$\overline{B}_{1.5}\equiv\overline{B}/(10^{1.5}\:{\rm \mu G})$ and
$\overline{n}_{H3}\equiv\overline{n}_H /10^{3}\:{\rm cm^{-3}}$. For
the diffuse ISM (Elmegreen 1985; Mestel 1985) with $\overline{n}_H
\sim 1\:{\rm cm^{-3}}$ and $\overline{B}\sim 3\:{\rm \mu G}$, we have
$M_B\sim 5\times10^{5}\:{\rm M_{\odot}}$, which is typical for a
GMC. Below this mass gravitational collapse is impossible without
ambipolar diffusion, and even for greater masses, there will still be
a significant effect. Collapse may also be impeded by turbulent
magnetic pressure, generated from energy injected by the first stars
to form in a cloud. McKee (1999) has modeled these higher mass clouds
as being in approximate hydrostatic equilibrium with low mass star
formation providing support. The data supporting the empirical Schmidt
laws (equations \ref{kenlaw} and \ref{larson}) are sensitive only to
high mass stars. Since collapse mediated by ambipolar diffusion occurs
on a timescale, $t_{AD}$, much greater than the free-fall time,
$t_{ff}$, we conclude it is inaccurate to use the rate of purely
gravitational perturbation growth of the disk averaged ISM and
ignoring magnetic fields, to predict global SFRs.

The spatial correlation of star formation with large scale spiral
structure in some disk galaxies motivates theories for the triggering
of star formation during the passage of gas through density
waves. Wyse (1986) and Wyse \& Silk (1989) propose a SFR law of the
form
\begin{equation}
\label{wyse}
\Sigma_{SFR}\propto \Sigma_{gas}^{N}(\Omega-\Omega_{p})
\end{equation}
where $\Omega_{p}$ is the pattern frequency of the spiral density
wave. In the limit of small $\Omega_{p}$ and for $N=1$ we recover
equation (\ref{larson}). The outer radius of star formation is
predicted to be the co-rotation radius, in approximate agreement with
observations. Increased cloud collision rates and increased
perturbation growth rates in the arms, where $Q$ is locally lowered,
have been suggested as the star formation triggering mechanism. These
will be further discussed in \S\ref{sec:spiral}. Ho, Filippenko \&
Sargent (1997) investigate the influence of bar density waves on star
formation in the nuclear regions of disk galaxies, finding
enhancements in the SFRs of early type spirals. They argue this is due
to the bar channeling gas to the central region, however the presence
of a bar is neither a necessary nor sufficient condition for nuclear
star formation. The wide range in the strength of nuclear HII regions,
whether a bar is present or not, suggests that it is not the passage
of gas through a density wave which mediates the star formation rate
within a particular starburst.

One prediction of these theories is a correlation of SFR with the
density wave amplitude. However this is not observed (Elmegreen \&
Elmegreen 1986; McCall \& Schmidt 1986, K89). Furthermore such
theories have difficulty explaining star formation in galaxies where
there is a lack of organized star formation features, as in flocculent
spirals (Block et al 1994; Sakamoto 1996; Thornley \& Mundy 1997a and
1997b; Grosbol \& Patsis 1998), some of which even contain moderate
amplitude density waves as revealed in the near infra-red. GMCs are
present and SFRs are similar to those systems where star formation is
organized into spiral patterns. This suggests stellar disk
instabilities, which create spiral density waves, and gas
instabilities, which lead to GMCs and large-scale star formation, are
decoupled in the sense that one does not cause the other (K89; Seiden
\& Schulman 1990). This decoupling highlights the need for a theory
which physically motivates the Schmidt Law of equation (\ref{larson})
without the need for coherent density waves.

In this paper we outline such a theory. The SFR, dominated by stars
forming in clustered regions, with high mass stars present, is
controlled by collisions between gravitationally bound gas clouds,
which can be atomic, molecular or both. We find the collision time is
a fraction of the orbital period. Collisions create localized,
over-dense regions where high mass star formation occurs. The
pre-collision clouds are formed relatively quickly by the action of
gravitational, thermal or Parker instabilities, growing in regions
where $Q\lesssim 1$. However, their collapse is halted by static and
turbulent magnetic pressure support. The latter may be produced by low
mass star formation regulated by ambipolar diffusion (e.g. McKee
1989), which does not dominate the global galactic SFR. Thus the rate
limiting step for star formation is not the formation of bound clouds,
but the compression of these, or parts of these, in cloud-cloud
collisions. Therefore at any particular time, most of the bound gas is
not undergoing collision induced star formation. There is no specific
need for large scale, coherent density waves. 

There is some evidence for collision induced star formation in the
Galaxy. Scoville, Sanders \& Clemens (1986) noted the efficiency per
unit mass of H$_2$ for OB star formation decreases significantly with
increasing cloud mass over the range $10^5$ to $3\times 10^6\:{\rm
M_{\odot}}$ (see also \S\ref{eff}) and concluded the principal trigger
for star formation is not an internal mechanism, such as the growth
rate of gravitational instability or sequential star
formation. Scoville et al suggested the approximately quadratic
dependence of the Galactic H~II region distribution on the local H$_2$
density (averaged on scales $\sim 300$ pc) was evidence for cloud
collisions causing massive star formation. Detailed observations of
individual star forming regions also suggest cloud collisions are an
important triggering mechanism (Scoville et al 1986; Maddalena et al
1986; Hasegawa et al 1994; Greaves \& White 1991; Womack, Ziurys, \&
Sage 1993). Clouds with embedded clusters of star formation have
broader distributions of optical polarization angles than clouds
without star formation (Myers \& Goodman 1991). This may indicate
accumulation of gas at super-Alfvenic speeds in these star forming
regions from the collision of two clouds with distinct magnetic field
alignments. Note, the results of Myers \& Goodman suggest the enhanced
dispersion in polarization angles is more closely associated with the
presence of dense gas than with the young stars which subsequently
form. Computer simulations of collisions between inhomogeneous clouds
(Klein \& Woods 1998) reveal the formation of high density clumps
embedded in filamentary structures, via a bending mode
instability. Such structures are abundant in OMC-1 (Wiseman \& Ho
1994, 1996) and Taurus (Ungerechts \& Thaddeus 1987). In modeling
starbursting systems IC 1908 and NGC 6872, Mihos, Bothun \& Richstone
(1993) were unable to reproduce observations of enhanced SFRs in
regions of high velocity dispersion and circular velocity gradients,
where cloud collision rates are high, with traditional,
non-collisional prescriptions of star formation.

While there is much evidence for cloud collisions playing an important
role in inducing star formation, it is not yet clear if the majority
of star formation is triggered by this process. The data supporting
equations (\ref{kenlaw}) and (\ref{larson}) are sensitive only to high
mass star formation, although the bulk of stars are expected to form
in these regions (Lada et al 1993). The theory of collision induced
star formation outlined below, requires the initial trigger for most
star formation to be a cloud collision. However, subsequent triggering
by other processes, such as self-propagating star formation, from the
initial site, is not excluded.

The rest of this paper is set out as follows. In \S\ref{setup} we set
out our preliminary assumptions, and derive results for gas disks,
independent of the hypothesis of collision induced star formation. In
\S\ref{sec:SFR} we derive a SFR law of the form $\Sigma_{SFR}\propto
\Sigma_{gas} \Omega$ in the case of uniform circular velocity, based
on star formation from cloud collisions. In \S\ref{testing} we compare
this law to observations and make predictions of radial SFR profiles,
SFR fluctuations due to different shear velocities, dependencies of
disk averaged SFR with gas density (Schmidt law) and circular velocity
(B-band Tully-Fisher relation) and how the efficiency of star
formation depends on cloud mass. Such tests are required to
distinguish this theory from those involving the growth rate of
gravitational perturbations or triggering by density waves. In
\S\ref{sec:spiral} we examine how large scale spiral density waves
affect the collisional theory. Finally, in \S\ref{sec:sb} we consider
the theory's application to the circumnuclear disks of starbursts,
which make up most of the dynamic range in the data supporting a
global Schmidt law (K98).

\section{Star Formation from Cloud Collisions}

\subsection{Preliminary Assumptions\label{setup}}

The star forming regions under consideration are thin,
self-gravitating disks. Self-regulated star formation (e.g. Silk 1997)
enforces the condition $Q\sim {\cal O}(1)$. The circumnuclear disks of
starbursts (DS98) and the star-forming regions of disk galaxies (K89)
satisfy these conditions. For $Q\lesssim 1$, an overdensity on scales
in the critical range leads to a bound object and we assume
instabilities drive most of the gas mass into bound clouds.
Gravitational, Parker and thermal instabilities have been considered
(e.g. Wada \& Norman 1999; Burkert \& Lin 1999; Elmegreen 1991). The
effect of cloud growth via collisional coagulation has also been
examined (e.g. Oort 1954; Field \& Saslaw 1965; Kwan \& Valdes 1987;
Das \& Jog 1996). Gas in the bound clouds can be either atomic or
molecular. In the Milky Way, extended HI envelopes are commonly
observed around molecular clouds (e.g. Moriarty-Schieven, Andersson \&
Wannier 1997; Williams \& Maddalena 1996; Elmegreen 1993; Blitz \&
Williams 1999, \S2.3). Although Andersson \& Wannier (1993) conclude
the HI around low mass ($\sim 10^{3}-10^{4} \:{\rm M_{\odot}}$) clouds
is not gravitationally bound, for larger GMCs ($M_c \sim 5\times
10^{5} \:{\rm M_{\odot}}$), the situation is probably reversed (Blitz,
private communication). In the solar neighborhood the mass in atomic
envelopes is similar to the mass in GMCs (Blitz 1990). Thus, out to
about 8 or 9 kpc, most of the gas in the Galaxy is in self-gravitating
clouds. As gas densities and pressures increase towards the centers of
galaxies, the molecular fraction of the gas is expected to increase,
until it is almost completely molecular in the circumnuclear disks of
starbursts (e.g. Liszt \& Burton 1996, DS98).

For simplicity we describe the cloud population with a single, typical
mass, $M_{c}$. In galactic disks, this approximation is justified by
the observed mass spectrum of GMCs in the Milky Way, $d{\cal
N}/dM\propto M^{-\beta}$, with $\beta\sim1.6$, and with an exponential
cutoff above $M_{cut}\sim 5\times 10^{6}\: {\rm M_{\odot}}$
(e.g. Solomon et al 1987). Note, these are only the molecular
masses. Most of the gas mass is in the large clouds. Observations
suggest circumnuclear disks are clumpy and the typical mass is much
larger (DS98). However, there is little evidence for the form of the
mass function. For galactic disks we set $M_c=5\times 10^{5}\: {\rm
M_{\odot}}$, while for circumnuclear disks we consider clouds with
$M_c=10^{8}\: {\rm M_{\odot}}$. The properties and timescales
associated with these clouds are shown in Tables (\ref{tab:prop}) and
(\ref{tab:times}).


The cloud radius, $r_{c}$, is smaller than its tidal radius, $r_{t}$,
defined as the radial distance from the cloud's center, at which the
shear velocity, $v_s$, due to differential galactic rotation, is equal
to the escape velocity from the cloud at that distance. The shear
velocity of two orbits separated by a radial distance, $b$, is
\begin{equation}
\label{v_s}
v_s(b)=b\left(\Omega-\frac{dv_{circ}}{dR}\right),
\end{equation}
and so for $b=r_t$
\begin{equation}
\label{r_tgeneral}
r_{t}= ( 1 - \beta)^{-2/3}\left( \frac{2 M_{c}}{M_{gal}}\right)^{1/3}\:R.
\end{equation}
$M_{gal}$ is the galactic mass interior to $R$, assuming a spherical
distribution. This approximation is valid at larger $R$, when the dark
matter halo begins to dominate over the disk mass. For smaller $R$,
and particularly for the circumnuclear disks of starbursts, this is
not the case. However, for simplicity, we keep this formalism, where
$M_{gal}$ is understood to be the ``equivalent interior galactic
mass'', if the distribution was spherical instead of
disk-like. Equation (\ref{r_tgeneral}) implies $r_t\rightarrow \infty$
for solid body rotation, when $dv_{circ}/dR=v_{circ}/R$ and thus
$\beta=1$. For most of the star-forming circumnuclear and galactic
disks rotation curves are flat (DS98; K89), $\beta=0$ and we have
\begin{equation}
\label{r_t}
r_{t}= \left( \frac{2 M_{c}}{M_{gal}}\right)^{1/3}\:R.
\end{equation}
$r_t$ is of order 100 pc for the fiducial values of $M_c$ in
circumnuclear and galactic disks. The relation between $r_c$ and
$r_t$ will be discussed in more detail in \S\ref{sec:SFR}.

The dimensions of the clouds are comparable to the scaleheight of the
gas disk (e.g. Solomon et al 1987; DS98), and so we describe the cloud
distribution with a thin, two dimensional disk. We also assume an
approximately axisymmetric distribution, so there is a single value of
$Q$ at any particular $R$. Galaxies with strong spiral arms and thus
non-axisymmetric gas distributions will be discussed in
\S\ref{sec:spiral}.

We assume the cloud velocity dispersion, $\sigma_{gas}$, results from
a balance of heating via gravitational torquing from non-collisional
encounters and cooling via dissipative collisions. Gammie et al (1991)
numerically integrated orbits for two-body encounters to obtain
\begin{equation}
\label{gammie}
\sigma_{gas}\simeq(GM_c\kappa)^{1/3}(1.0-1.7\beta),
\end{equation}
valid for $\beta \ll 1$ and in approximate agreement with Galactic
observations (e.g. Stark \& Brand 1989; Knapp, Stark \& Wilson 1985;
Clemens 1985). For a flat rotation curve, this is approximately the
shear velocity of an encounter of impact parameter, $b=r_t$. The
surface densities of real disks, set by $Q\sim {\cal O}(1)$, are such
that the effects of many-body interactions may be important. N-body
simulations are required to probe these effects. Substituting for
$\sigma_{gas}$ in equation (\ref{Q}), we derive the radial
distribution of gas,
\begin{equation}
\label{radialgas}
\Sigma_{gas}=\frac{\alpha \kappa^{4/3} M_c^{1/3}}{\pi G^{2/3}Q}\propto M_c^{1/3}\left(\frac{v_{circ}}{R}\right)^{4/3}\frac{(1+\beta)^{2/3}(1.0-1.7\beta)}{Q}\propto \frac{M_c^{1/3}v_{circ}^{4/3}(1-1.0\beta))}{R^{4/3} Q}.
\end{equation}
Note K89, assumed $\sigma_{gas}$ was independent of $R$, which leads
to an underestimation of $Q$, by factors of a few, in the central
galactic regions compared to the case where equation (\ref{gammie}) is
used instead. This may explain the slight trend of $Q$ decreasing by
factors of a few as one moves towards the centers of galaxies, (K89,
figure 11), rather than remaining constant. However, better statistics
are required before a proper comparison can be made.

When $Q\gg 1$, the assumption that most of the gas mass is organized
in bound clouds breaks down, together with our use of equation
(\ref{gammie}). The presence of a large scale stellar bar, channeling
gas radially inwards, will deplete the gas from certain regions, thus
raising $Q$. Here we expect little or no star formation.  This may be
the situation in the inner few kpc of the Milky Way (e.g. Binney et al
1991).

\subsection{The Collision Induced SFR \label{sec:SFR}}

Our principal hypothesis is that cloud collisions, by compressing
parts of the clouds, induce the majority of star formation in galactic
and circumnuclear disks. However, collisions can also be disruptive.
A simple theoretical condition for colliding clouds to remain bound
has been given by Larson (1988). Neglecting post-shock cooling, the
clouds stay bound if the ram pressure, $\sim \rho_{c} v_{rel}^{2}$, is
less than the binding pressure, $\sim G \Sigma_{c}^{2}$, where
$\rho_c$ and $\Sigma_c$ are the cloud volume and surface densities
respectively.  For typical galactic disk cloud properties in table
\ref{tab:prop}, this implies $\Sigma_c\gtrsim 900 \:{\rm
M_{\odot}\:pc^{-2}}$, which is higher than the mean value for GMCs
($\sim 170 \:{\rm M_{\odot}\:pc^{-2}}$) by a factor of about
five. Realistic clouds are probably more extended, with a gradually
decreasing density profile. Interactions in these outer layers reduce
the actual relative velocity from the value quoted in
table~\ref{tab:prop}. Radiative post-shock cooling reduces the
disrupting pressure, and thus also relaxes the above
condition. Therefore we expect some collisions to lead to an increase
in mass, density and the gravitational potential energy of the clouds
involved, and hence the likelihood of faster SFRs.

The outcome of cloud collisions has also been investigated numerically
(e.g. Lattanzio et al 1985), but the simulations have usually been
unable to resolve the Jeans length, thus violating the numerical Jeans
condition (Truelove et al 1997). Furthermore, there has been no
systematic attempt to probe the parameter space of cloud collisions,
as defined by the angle of collision, impact parameter, Mach number
and mass ratio. Magnetic fields have yet to be included. For the
idealized cases considered, the results of a collision depend
sensitively on the collision parameters (Klein, private communication;
Lattanzio et al 1985). We make the simplifying assumption that the
fraction of a cloud converted into stars in a typical collision,
averaging over the parameter space of possible collisions, is
constant.

We consider the thin disk of self-gravitating clouds described in
\S\ref{setup}, where $Q\sim {\cal O}(1)$. We hypothesize
$\Sigma_{SFR}$ is, on average, inversely proportional to the collision
time, $t_{coll}$, of these clouds. A fraction, $\epsilon$, of each gas
cloud is converted into stars in each burst of collision induced star
formation. The time between bursts is $f_{sf}^{-1} t_{coll}$, where
$f_{sf}$ is the fraction of collisions which lead to star
formation. Thus,
\begin{equation}
\label{sfr1}
\Sigma_{SFR}=\frac{\epsilon f_{sf} {\cal N}_A M_{c}}{t_{coll}} \simeq \frac{\epsilon f_{sf} \Sigma_{gas}}{t_{coll}},
\end{equation}
where ${\cal N}_{A}$ is the surface number density of gravitationally
bound clouds per unit area of the disk. By numerically solving the
equations of motion, Gammie et al (1991, figure 8), found cloud-cloud
collisions result from encounters caused by differential rotation,
primarily with initial impact parameters of about $1.6 r_{t}$, and
with a spread in values of order $r_t$\footnote{The length unit used
in Gammie et al (1991) corresponds to $\sim0.8 r_t$}. For typical GMC
parameters in the Galaxy, the associated shear velocity is $\sim 9
\:{\rm km/s}$. This sets the collision rate, together with the cloud
surface density, ${\cal N}_A$, and the probability of collision,
$f_G$, of these encounters. Note, the random velocity dispersion of
the cloud population ($\sim 7\:{\rm km/s}$ e.g. Stark \& Brand 1989)
sets the clouds moving on epicycles, but is not the velocity directly
influencing the collision rate. The effect of these random motions has
been accounted for in the calculations of Gammie et al, since they
consider the collision of clouds which are already moving on
epicycles. Increasing the random motions increases the initial impact
parameters at which most cloud collisions occur, raising the shear
velocity and thus the collision rate. We express $t_{coll}$ as
\begin{equation}
\label{taucoll}
t_{coll}\sim\frac{1}{2}\frac{\lambda_{mfp}}{v_{s}(\sim1.6 r_{t})}\sim\frac{1}{3.2 r_{t}(\Omega -\frac{dv_{circ}}{dR}){\cal N}_{A} r_{t}f_G},
\end{equation}
where the first factor of $1/2$ accounts for clouds either catching up
with others at larger $R$ or being caught up with by clouds at smaller
$R$. $\lambda_{mfp}=1/{\cal N}_A r_t f_G$ is the mean free path of a
cloud to catch up, or be caught up to, by another. $v_{s}(\sim1.6
r_{t})\simeq 1.6 r_{t}(\Omega-\frac{dv_{circ}}{dR})$ is the shear
velocity of an encounter with impact parameter $\sim 1.6 r_{t}$, due to
differential rotation. 

We evaluate the factor ${\cal N}_A r^{2}_{t}$ via
\begin{equation}
\label{N}
{\cal N}_A\simeq\frac{\Sigma_{gas}}{M_{c}}=\frac{\alpha \kappa \sigma_{gas}}{\pi G Q M_c}\simeq(1+0.3\beta)\frac{0.7\alpha}{Qr_{t}^{2}}.
\end{equation}
As in equation (\ref{radialgas}), we have used $\kappa=
\sqrt{2}\Omega(1+\beta)^{1/2}$ and assumed the velocity dispersion of
the gas clouds results from gravitational torquing (Gammie et al 1991)
so that $\sigma_{gas}\simeq (G M_c \kappa)^{4/3}(1.0-1.7\beta)$, with
$\beta\ll 1$. So ${\cal N}_A \pi r^{2}_{t}= (1+0.3\beta)0.7 \alpha \pi
/ Q \sim {\cal O}(1)$ and is constant where $Q$ is constant. Thus
every area element, $\pi r_{t}^{2}$, of the disk approximately
contains the mass of gas, $M_c$, required to set $r_{t}$. Thus, from
equation (\ref{taucoll}),
\begin{equation}
\label{colltime}
t_{coll}\simeq \frac{Q}{9.4 f_G (1+0.3\beta) (1-\beta)} t_{orb}.
\end{equation}

From Gammie et al (1991) we set $f_G\sim0.5$. We expect it to scale as
$r_c/r_t$. We consider cloud boundaries to be set by pressure
confinement from the general ISM pressure, $P_{ISM}$. Following
Elmegreen (1989) we have
\begin{equation}
\label{pism}
P_{ISM}\simeq \frac{\pi}{2}G\Sigma_{gas}\left(\Sigma_{gas}+\Sigma_{*}\frac{\sigma_{gas}}{\sigma_{*}}\right),
\end{equation}
where $\Sigma_*$ and $\sigma_*$ are the stellar surface density and
velocity dispersion respectively. The boundary pressure of the
self-gravitating clouds is a few times less than the interior cloud
pressure, $P\sim\frac{1}{2}G\Sigma_{c}^{2}$, where $\Sigma_c\simeq M_c
/\pi r_c^{2}$. Since $Q\sim{\cal O}(1)$ implies $\Sigma_{gas}\simeq
M_c / (\pi r_t^2)$, and with $P\sim P_{ISM}$, we have
\begin{equation}
\label{rcrt}
\frac{r_c}{r_t}=\left(\frac{\Sigma_{gas}}{\Sigma_c}\right)^{1/2}\sim\left(\frac{\Sigma_{gas}}{(\Sigma_{gas}+\Sigma_{*}\frac{\sigma_{gas}}{\sigma_{*}})}\right)^{1/4}.
\end{equation}
Observationally, $\Sigma_{gas}$ and $\Sigma_*$ have approximately
similar spatial distributions, and so from equation (\ref{rcrt}) we
see that $r_c / r_t$, and thus $f_G$, varies only very slowly with
$R$. From here on we take it to be a constant.

Substituting equation (\ref{colltime}) into equation (\ref{sfr1}), we
obtain
\begin{equation}
\label{sfr3}
\Sigma_{SFR}\simeq 1.5 \epsilon f_{sf} f_G Q^{-1} \Sigma_{gas} \Omega (1-0.7 \beta).
\end{equation}
This is a new ``modified''-Schmidt Law, to be tested against
observations (\S\ref{testing}). For our fiducial location in the
Galactic disk ($R=4$ kpc) we have
\begin{equation}
\label{sfr4}
\Sigma_{SFR}\simeq 4.3\times 10^{-8}\:{\rm  M_{\odot}\:yr^{-1}\:pc^{-2}}\:\left( \frac{\epsilon}{0.2} \frac{f_{sf}}{0.5} \frac{f_G}{0.5} \frac{1.0}{Q} \frac{\Sigma_{gas}}{10\:{\rm M_{\odot}\:pc^{-2}}} \frac{\Omega}{5.7\times 10^{-8}\:{\rm yr^{-1}}} (1-0.7 \beta)\right).
\end{equation}
Disk averaged SFRs, with the appropriate gas distribution, are
estimated in \S\ref{diskav}.

\subsection{Predictions of Collision Induced Star Formation\label{testing}}

\subsubsection{Radial Profiles\label{mihos}}

With high resolution data for $\Sigma_{SFR}$, $\Sigma_{gas}$,
including atomic and molecular components, and $v_{circ}$, equation
(\ref{sfr3}) can be directly tested. This is practical for the Milky
Way and nearby galaxies, but difficult for circumnuclear disks of
starbursts because of their small size. Star formation from cloud
collisions is a stochastic process and so statistically significant
data sets are required. Properly identifying bound clouds requires
atomic and molecular observations, so the masses of both components
can be accounted for.

The assumption that the cloud velocity dispersion is caused by
gravitational torquing (Gammie et al 1991), also leads to the
prediction of $\Sigma_{gas}(R)$ (equation \ref{radialgas}). Combining
this with equation (\ref{sfr3}) leads to
\begin{equation}
\label{sfrr}
\Sigma_{SFR}(R)\propto M_c^{1/3} \Omega^{7/3} Q^{-2} (1-1.7\beta),
\end{equation}
which is proportional to $M_c^{1/3}R^{-7/3}Q^{-2}$ for constant
$v_{circ}$. If observations of $\Sigma_{gas}$ are lacking, then the
theory can still be tested using equation (\ref{sfrr}) and SFR and
circular velocity data, for an assumed constant $Q$. Note, $M_{c}(R)$
is, in general, difficult to determine. However surveys of Galactic CO
(e.g. Sanders et al 1986) find no strong evidence for systematic
variation (Solomon et al 1987; Scoville et al 1987). Furthermore any
variation is weakened by being raised to the $1/3$ power in equation
(\ref{sfrr}). If galactic stellar disks have been built up primarily
through self-regulated star formation, where $Q\sim {\cal O}(1)$, then
we also have $\Sigma_{*}\propto \Sigma_{SFR}$ as an additional
prediction.

Several authors have presented radial profiles of $\Sigma_{gas}$ and
$\Sigma_{SFR}$ for individual galaxies (e.g. Tacconni \& Young 1986;
Kuno et al 1995). However, problems of accounting for the varying
extinction of the tracers of star formation, such as H$\alpha$ and
Br$\gamma$ make direct comparison difficult. Similarly, where FIR
emission is used as a SFR estimator, the heating contributions from
young stars, old stars and possible AGN activity must be disentangled.
A follow-up paper to K98, (Martin \& Kennicutt 2000), will present
radial data for many galaxies, accounting for these effects. 

One distinct prediction of this theory results from the extra
dependence of the SFR on variations in the circular velocity.
Statistically, we expect negative velocity gradients in the rotation
curve to increase the SFR and positive gradients to decrease
it. Regions of solid body rotation will be free of collisions
resulting from shearing motions. Thus we expect star formation here to
have a different triggering mechanism. These regions provide a good
control environment for testing the collisional theory. In general, we
expect a positive correlation between the SFR and the velocity
dispersion of the clouds. With higher random velocities they move on
larger epicycles, and encounters with greater initial impact
parameters and greater shear velocities occur, leading to increased
collision rates\footnote{Complications arise if the likelihood of star
formation changes significantly with the increase in relative velocity
of the collision. Numerical simulations are required to investigate
this effect.}. No such prediction is made by the theory of star
formation triggered by the rate of gas passage through spiral arms. In
a future paper we plan to investigate this question with high
resolution BIMA and VLA data.

\subsubsection{Disk Averages - Schmidt Law and Tully-Fisher Relation\label{diskav}}

We also test equation (\ref{sfr3}) by examining the disk averaged
properties of galaxies and starbursts. We note, however, that such
tests, while good consistency checks, do not discriminate well between
the different theories of how star formation is triggered. We take the
area-weighted mean of equation (\ref{sfr3}) over the whole region of a
disk, where $Q\sim 1$ and $v_{circ}$ is constant, to obtain
\begin{equation}
\label{sfr5}
\overline{\Sigma_{SFR}}\equiv\frac{1}{\pi(R_2^2-R_1^2)}\int_{R_1}^{R_2}\Sigma_{SFR}(R)2\pi R\:dR=1.5\epsilon f_{sf} f_{G}\overline{\Sigma_{gas} \Omega}.
\end{equation}
Current observations do not have the spatial resolution to estimate
$\overline{\Sigma_{gas}\Omega}$, except for the nearest galaxies.
However, K98 presents data revealing a correlation between
$\overline{\Sigma_{SFR}}$ and $\overline{\Sigma_{gas}}\:\Omega_2$,
where $\Omega_2$ is the angular rotation frequency at the outer
radius, $R_2$, of the star forming region (figure
\ref{fig:ken2}). Since we are considering the flat rotation curve
case, we rewrite equation (\ref{sfr5}) as
\begin{equation}
\label{sfr6}
\overline{\Sigma_{SFR}}=1.5\epsilon f_{sf} f_{G} \Omega_{2} R_{2} \overline{\left(\frac{\Sigma_{gas}}{R}\right)}.
\end{equation}
For $\Sigma_{gas}\propto R^{-4/3}$ (equation \ref{radialgas}) we obtain
\begin{equation}
\label{sfr7}
\overline{\Sigma_{SFR}}=3\epsilon f_{sf} f_{G}\Omega_{2}\overline{\Sigma_{gas}}\left(\frac{x^{-1/3}-1}{1-x^{2/3}}\right),
\end{equation}
where $x=R_1/R_2$, $R_1$ being the inner radius where the rotation
curve is flat. If $x$ is uncorrelated with $\Omega_{2}$ and
$\overline{\Sigma_{gas}}$, then we predict K98's observed correlation.

Applying equation (\ref{sfr7}) to the inner 8.5 kpc of the Milky Way,
assuming $x=0.2$ and $\epsilon=0.2$ (see \S\ref{eff}), gives
\begin{equation}
\label{milkyway}
SFR_{tot}=\overline{\Sigma_{SFR}}\pi R_2^2= 1.6\:{\rm M_{\odot}\:yr^{-1}}\:\left(\frac{\epsilon}{0.2} \frac{f_{sf}}{0.5} \frac{f_G}{0.5} \frac{1.0}{Q} \frac{R_2}{8.5\:{\rm kpc}}\frac{v_{circ}}{225\:{\rm km/s}} \frac{\overline{\Sigma_{gas}}}{10\:{\rm M_{\odot}\:pc^{-2}}}\right).
\end{equation}
This is consistent with estimates based on observations of thermal
radio emission from HII regions, which give $SFR_{tot}=2.7\pm0.9\:{\rm
M_{\odot}\:yr^{-1}}$ (G\"usten \& Mezger 1982; McKee 1989). Scaling to
a typical starbursting circumnuclear disk, with $R_2=1.7\:{\rm kpc}$,
$v_{circ}=300\:{\rm km/s}$ and $\overline{\Sigma_{gas}}=10^3\:{\rm
M_{\odot}\:pc^{-2}}$, gives $SFR_{tot}=42\:{\rm
M_{\odot}\:yr^{-1}}$. One of the most uncertain factors in these
estimates is $f_{sf}$. Numerical studies, (Klein, private
communication), although not yet incorporating magnetic fields, can
constrain this number.

We also predict the form of the B band Tully-Fisher relation, $L_B
\propto v_{circ}^{\alpha_{TF}}$. Assuming $L_B \propto
\overline{\Sigma_{SFR}} R_2^2$ we have
\begin{equation}
\label{tully}
L_B \propto v_{circ} R_2 \overline{\Sigma_{gas}} \propto v_{circ}^{7/3} R_2^{-1/3} M_c^{1/3},
\end{equation}
where we have evaluated $\overline{\Sigma_{gas}}$ using equation
(\ref{radialgas}) assuming $R_1 \ll R_2$. This result compares
favorably with the observed B band exponent of $\alpha_{TF}\simeq 2.1
- 2.2$ (Burstein et al 1995; Strauss \& Willick 1995). The
Tully-Fisher relation at longer wavelengths becomes more and more
contaminated by light from older stellar populations.

\subsubsection{Cloud Star Formation Efficiency\label{eff}}

Collision induced star formation predicts a variation in the star
formation efficiency, $\epsilon$, of GMCs, dependent on their
mass. Variations in $\epsilon$, resulting from this and other
mechanisms, have been considered by a number of authors
(e.g. Elmegreen \& Clemens 1985; Scoville et al (1986); Pandey,
Paliwal \& Mahra 1990; Franco, Shore \& Tenorio-Tagle 1994; Ikuta \&
Sofue 1997; Williams \& McKee 1997).

Scoville et al (1986) found the ratios of Lyman-continuum luminosity
to $M_{c}$ and number of high luminosity HII regions to $M_{c}$,
decreased with increasing $M_c$. They argued this was evidence for
collision induced star formation, since, if the collision rate scaled
as the cloud surface area ($\propto M_c^{2/3}$), then the efficiency
of star formation per unit cloud mass, $\epsilon$, would scale as
$M_c^{-1/3}$. However, if a more appropriate mass-size relationship
($M_c \propto r_c^{2}$; Larson 1981) is applied with their reasoning,
then no scaling of $\epsilon$ with $M_c$ is predicted.

Ikuta \& Sofue (1997) considered the (radio) luminosity of HII region,
rather than simply the number, associated ($<10$ pc away in the plane
of the sky) with molecular clouds. They found $\epsilon \propto
M_c^{-0.78}$. However, they did not allow for higher mass clouds being
larger than smaller ones. The centers of large clouds may be much
further than 10 pc away from HII regions associated with their
periphery. There is also no comment on the completeness of the
data. In particular the HII region sample, being flux limited at
$1\:{\rm Jy}$, is incomplete below luminosities of $\sim 250\:{\rm
Jy\:kpc^{2}}$ for sources in the inner Galaxy\footnote{A luminosity of
$1\:{\rm Jy\: kpc^{2}\simeq 0.012\:{\rm S_{49}}}$, where $S_{49}$ is
the ionizing luminosity ($\lambda<912\:{\rm \AA}$) in units of
$10^{49}\:{\rm photons \:s^{-1}}$. For example, Orion A has
$S_{49}=2.7$ and $L=230\:{\rm Jy\:kpc^{2}}$. Note, by definition an
object with a radio flux of 1 Jy at a distance of 1 kpc has a
luminosity of $1\:{\rm Jy\:kpc^{2}}$, i.e. there is no factor of
$4\pi$.}.

We now revisit the question of how the efficiency, $\epsilon\equiv
M_{*}/(M_* + M_c) \simeq M_*/M_c$, of star formation depends on cloud
mass. Observationally, we measure $\epsilon$ by summing the radio
luminosity, $L_{rad}$, of HII regions associated with a cloud. This
luminosity is converted into a stellar mass using $M_*=570
S_{49}\:{\rm M_{\odot}}=6.84 L_{rad}\:{\rm M_{\odot}\:Jy^{-1}\:kpc^2}$
(McKee \& Williams 1997). Dividing by $M_c$ then gives $\epsilon$. We
use the HII region data of Downes et al (1980) and the molecular cloud
data of Solomon et al (1987). For each of the 106 HII regions, with
resolved distance ambiguity, in the same region as the cloud survey,
molecular clouds within $40\:(M_c/5\times 10^{5}\:{\rm
M_{\odot}})^{1/2}$ pc (i.e. 2 $r_c$) on the sky, within the correct
radial distance bounds and with relative velocities of $<15\:{\rm
km/s}$ are identified. The HII region is associated with the closest
cloud, if more than one is identified. With these criteria, 83 HII
regions are associated with 39 clouds. This data is shown in figure
\ref{fig:eff}a, with the typical errors shown by the cross.

However, there are also two issues of completeness for the
sample. Firstly, as already mentioned, the HII data are incomplete for
$L<250\:{\rm Jy\: kpc^{2}}$. We adopt the more conservative limit of
$400\:{\rm Jy\: kpc^{2}}$, which is shown by the diagonal dashed line
in figure \ref{fig:eff}a. The data below this line are
incomplete. Secondly, the molecular cloud data are incomplete for
$M_c\lesssim 4\times 10^{5}\:{\rm M_{\odot}}$ (Williams \& McKee
1997). This is shown by the vertical dashed line. Assuming, for a
particular $M_c$, the probability of detection of a cloud is
uncorrelated with $L_{rad}$, we ignore the effect of cloud
incompleteness in the following analysis. The total $L_{rad}$ for each
cloud is recalculated, now only summing individual HII regions with
$L_{rad}>400\:{\rm Jy\: kpc^{2}}$. This data is shown in figure
\ref{fig:eff}b. The best fit straight line is shown by the long dashed
line.

We now compare these observations to theory. We model a cloud of mass
$M_c$, with HII regions, as resulting from a binary collision of two
smaller clouds of mass $M_1$ and $M_2$, with $M_1+M_2=M_c$ and
$M_1<M_2$. The assumption of binary collisions is justified by the
paucity of high mass star formation in the total cloud population.
Given our lack of understanding of the detailed results of cloud
collisions, the assumption that several HII regions, rather than just
one, may result from a single collision event is reasonable. $M_1$
(and $M_2$) are chosen from the mass spectrum of clouds: $d{\cal
N}/dM\propto M^{-\beta}$, with $\beta\sim1.6$. Following our original
hypothesis applied to equal mass clouds, that, on average, in a
collision a constant fraction, $\epsilon$, of the total cloud mass
forms stars, we assume for unequal mass collisions, the mass of stars
formed is $2 \epsilon M_1$. Thus $L_{rad}\propto M_1$, and the maximum
star formation efficiency, $\epsilon_{max}$, is statistically,
achieved for collisions of equal mass clouds ($M_1=M_c/2$), and is
thus independent of $M_c$. From the data in figure \ref{fig:eff}b, we
set $\epsilon_{max}=0.2$. The observationally determined minimum
luminosity, above which the HII region sample is complete, defines a
minimum value, $M_{min}$, for $M_1$. Thus $M_{min}<M_1<M_c /2$. These
two constraints meet when $M_{min}=M_c/2$, and thus the star formation
efficiency data for a complete sample of HII regions associated with
clouds will be contained within a triangular region in a diagram of
${\rm log}\:\epsilon$ vs. ${\rm log}\:M_c$ (figure \ref{fig:eff}b).

The typical value of $M_1$, $\left< M_1 \right>$, is an average of
$M_1$ weighted by the collision rate. This scales with $M_c$ via
\begin{equation}
\left< M_1 \right> =\frac{ \int_{M_{min}}^{M_c/2}M_1 t_{coll}^{-1}(M_1) \: dM_1}{\int_{M_{min}}^{M_c/2} t_{coll}^{-1}(M_1) \: dM_1}\propto M_c^{0.4},
\label{M1dist}
\end{equation}
in the region where $M_{min}\ll M_c$, since from equation
(\ref{taucoll}), we have
\begin{equation}
\label{rate2}
t_{coll}^{-1}(M_1)\propto {\cal N}^{}_{A,sum} r_{t,sum}^{2} \propto M_1^{-1.6},
\end{equation}
and since $r_{t,sum}=r_{t,1}+r_{t,2}\propto M_c^{1/3}$ is
approximately independent of $M_1$, and ${\cal N}_{A,sum}={\cal
N}_{A,1}+{\cal N}_{A,2}\simeq{\cal N}_{A,1}\propto M_1^{-1.6}$. Thus
$\left< M_1 \right>\propto M_c^{0.4}$ and so
\begin{equation}
\label{burst}
\left< \epsilon \right> = 2 \epsilon \left< M_1 \right>/M_c \propto M_c^{-0.6}.
\end{equation}
Performing an averaging of the distribution of $M_1$ (and thus
$\epsilon$) given by equation \ref{M1dist}, in logarithmic space,
leads to line A in figure \ref{fig:eff}b. 

We also consider three other theoretical models for comparison. Models
B and C (stochastic star formation, Williams \& McKee 1997) assume a
cloud's star formation rate is linearly proportional to its mass,
provided $\epsilon_{max}$ is not exceeded. Model B imposes a mass
dependence, at high masses, on $\epsilon_{max}$ by assuming there is a
physical limit to the maximum luminosity of OB associations. Model C
assumes $\epsilon_{max}$ independent of $M_c$, and the reason for the
observed maximum association luminosity is simply due to sparse
sampling of the distribution. By normalizing to the observed Galactic
OB association luminosity and molecular cloud mass functions, a
probability distribution of association luminosities expected in a
cloud of mass $M_c$, approximately proportional to $L^{-2}$ is
predicted. After summing the expected number of associations in a
cloud, and then averaging the resulting distribution of $\epsilon$ at
each $M_c$ in logarithmic space, we obtain lines B and C. Finally we
consider an unphysical model (D) with a uniform distribution of ${\rm
log}\:\epsilon$ for clouds of a particular mass. 

The four model predictions are shown in figures \ref{fig:eff}b
and~\ref{fig:eff}c. In the latter, we also show the 95\% confidence
limits\footnote{Based on assumption of normal distribution of data
about best fit line. The limits on acceptable slopes of fits are shown
by the asymptotic limit of these confidence limits. Fits with more
extreme slopes but still within the boundaries are also excluded.} on
the linear best fit from the existing sample (1), a hypothetical
sample ten times larger with the same distribution (2), and as for (2)
but with individual measurement errors reduced by a factor of two
(3). From the existing data, model D is marginally excluded (because
of its slope), while A,B,C are all consistent. Although the confidence
limits on fitting a non-linear function will differ in detail, figure
\ref{fig:eff}c qualitatively shows that with ten times more data and
with modest improvements in measurement accuracy, we can hope to
discriminate between these models. Such a sample can be achieved by a
survey of radio HII regions in the entire inner Galaxy to fluxes of
order $0.1\:{\rm Jy}$. Improving completeness of CO observations of
smaller mass GMCs will also improve the statistics. Uncertainties in
$M_c$ result from the assumption of virialization and the CO/${\rm
H_2}$ ratio. Errors in $L_{rad}$ result in part from distance
uncertainties. Direct infra red observations of massive stars may
improve estimates of association luminosities.

As a final point of caution, we note that evolutionary effects will
cause a cloud's measured $\epsilon$ to change. At the onset of star
formation $\epsilon$ will be small, while $\epsilon$ will increase as
the cloud is destroyed by energy injection from its high mass
stars. Large enough samples are required to average over this effect.

In summary, collision induced star formation predicts a decrease in
the mean cloud star formation efficiency of a complete sample with
increasing cloud mass. The decrease occurs since it is much more
likely for a large cloud to suffer a collision with a smaller one,
because of the cloud mass spectrum. It is the smaller cloud which then
determines the amount of resultant star formation. An observational
lower limit to HII region luminosities, included in the analysis,
implies a constant minimum mass of the smaller cloud, which drives the
mass dependence of $\epsilon$. Stochastic models predict a similar,
but less steep, decline due to the shape of the complete region in the
$\epsilon$ versus $M_c$ parameter space. For stochastic models, at
high cloud masses, if $\epsilon_{max}$ is independent of
$M_c$\footnote{A new physical regime may be reached at very high cloud
masses in the extreme conditions of circumnuclear starbursting disks,
when the ionized gas is no longer able to escape from the cloud. This
will have a profound effect on the efficiency of star formation. The
analysis of this section can thus only be applied to clouds in
``normal'' galactic environments.}, the efficiencies tend towards a
constant value. Improved data samples should allow for discrimination
between the models.

\subsection{Effects of Spiral Density Waves\label{sec:spiral}}

The theory for collision induced star formation requires modification
where there is a tight spatial correlation of star formation with
spiral structure. Spiral density waves decrease the local value of
$Q$, and often $Q<1$ in the arm region and $>1$ in the inter-arm
region (e.g. Kuno et al 1995). Two scenarios are possible. In the
first, the rate limiting step is the formation of bound clouds,
occurring exclusively in the arms, after which the clouds form stars
at a fast rate (e.g. via gravitational collapse of magnetically
supercritical clouds or through rapid collisions) and all the bound
gas is involved in star formation. Equation (\ref{wyse}) is then a
better description of the galactic SFR. In the second scenario, bound
cloud formation is fast and is not the rate limiting step. A reservoir
of bound gas clouds exists in the galaxy, including in the inter-arm
regions. Spiral arms now act to concentrate the spatial distribution
of gas clouds, and the collision rate is enhanced in the arms. The
individual collision rate for a particular cloud is still described by
equation (\ref{taucoll}), but the overall SFR is modulated by the
length of time the gas clouds spend in the arm region (related to
$\Omega-\Omega_p$ and the width and pitch angle of the arm) and the
degree of spatial concentration of clouds in the arm (related to the
strength of the spiral density wave).

In M51, which has strong, well-defined spiral arms, there is evidence
that the cloud collision times in the arm are short compared to the
arm crossing time (Kuno et al 1995), thus favoring the first
scenario. However, the similarity of the global star forming
properties of galaxies with and without large scale density waves
argues against bound cloud formation being controlled exclusively by
spiral arms. Deciding which scenario is the correct description, or if
both processes operate at some level, requires study of the arm to
inter-arm gas distributions and cloud collision timescales in larger
samples of galaxies, which exhibit a tight correlation of star
formation with spiral arms. Obviously, neither scenario can explain
the star formation observed in the galaxies without strong spiral
structure.

\subsection{Application to Circumnuclear Disks of Starbursts\label{sec:sb}}

The $\Sigma_{SFR}$ and $\Sigma_{gas}$ data of the circumnuclear disks
of starbursts make up most of the dynamic range for K98's Schmidt law
relationships (figures \ref{fig:ken1} and \ref{fig:ken2}). However,
much less is known about the details of star formation occurring in
these regions than in the outer regions of galactic disks.

Downes \& Solomon (1998, DS98) present observations and models of 10
circumnuclear disks. Their key findings, relevant to the theory of
collision induced star formation are the following: most of the
circumnuclear disks are in the flat rotation curve
regime\footnote{This includes most of the gas mass. Although the data
are limited, most of the star formation as traced by the ``extreme
starburst events'' (DS98, table 12) is also in this regime. The mean
rotation curve turn over radius is 220 pc, the mean half CO intensity
radius is 630 pc and the mean outer disk boundary (used in the disk
average analysis of K98) is 1.7 kpc.}; $Q\sim {\cal O}(1)$; the disks
are thin and modeled without need to invoke large-scale
non-axisymmetric features, such as bars or spiral arms; much of the
star formation is associated with very large bound gas clouds with
maximum masses $\sim10^{9}\:{\rm M_{\odot}}$ and sizes ($\sim 100$ pc)
consistent with their confinement by tidal shear forces. Thus our
principal assumptions are met.

The gas disks are predominantly molecular, including the inter-cloud
medium, with the gas mass approximately $1/6$ of the dynamical mass.
Note, larger gas masses (by a factor of $\sim5$) are derived if the
standard CO/H$_2$ conversion factor for virialized GMCs in normal
galaxies is used (as in the analysis of K98). In fact, much of the CO
luminosity of the circumnuclear disks comes from the non-virialized
molecular inter-cloud medium.

Since $Q\sim {\cal O}(1)$ and not $\ll 1$, the gas is probably not
collapsing on the free-fall timescale ($t_{ff}<10^{6}$ yrs). This is
consistent with Downes \& Solomon's estimate that about half the gas
is converted into stars over 10 orbital periods ($\sim 100 \times
10^{6}$ yrs). This efficiency per orbital period is similar to that of
normal galactic disks (K98), motivating a unified theory to describe
both regimes. The collision time for a typical cloud mass, which we
take to be of order $10^{8}\:{\rm M_{\odot}}$, is a few million years.

The one inconsistency between our theory and the results of DS98, is
their estimate that only $\sim 10$\% of the gas mass is in bound
clouds. They base this estimate on observations of HCN which requires
densities $\sim 10^{5}\:{\rm cm^{-3}}$ for excitation. However, the
fact that $Q\sim 1$ implies that any overdense region on the critical
scales will become gravitationally bound. A large fraction of the gas
may be at densities less than that required for strong HCN emission,
and yet still in bound clouds.

Circumnuclear starburst disks and the star forming regions of normal
galactic disks may represent the extremes of a continuous family of
states. Our theory of collision induced star formation can be applied
in both situations. As one moves inwards from the outer disk, the gas
surface densities and the molecular fractions of the bound clouds
increase, but self-regulation of star formation maintains $Q\sim {\cal
O}(1)$. The typical cloud mass, $M_c$ also appears to increase. If
this is set by the magnetic critical mass, $M_B$, of the inter-cloud
medium then, from equation (\ref{M_B}), $\overline{B}^{3}
/\overline{n}^{2}$ must increase as $R$ decreases. The typical cloud
mass may also be affected by changes in the collision time. This is
much shorter ($\sim 2.4 f_{sf}^{-1}\times 10^{6}$ yrs) in
circumnuclear disks than in the typical star-forming locations of
normal disks ($\sim 22 f_{sf}^{-1}\times 10^{6}$ yrs), particularly in
comparison to the stellar evolutionary and cloud destruction
timescales. This may increase $M_c$ in circumnuclear disks to be
several times $M_B$.

\section{Conclusions}

We have invoked star formation triggered by cloud-cloud collisions to
explain global star formation rates in disk galaxies and circumnuclear
starbursts. Previous theories based on the growth rate of
gravitational perturbations ignore the dynamically important presence
of magnetic fields. Theories based on triggering by spiral density
waves fail to explain star formation in systems without such
waves. Furthermore, observations suggest gas and stellar disk
instabilities are decoupled.

Star formation resulting from cloud collisions has been proposed in
the past (e.g. Scoville et al 1986), but rejected because of
supposedly long collision timescales. However, Gammie et al (1991)
show the collision rate of self-gravitating particles in a
differentially rotating disk is much larger than that of particles in
a box. Collision rates are enhanced because particles collide at the
shear velocity of encounters with initial impact parameters of order
two tidal radii (typically a few hundred parsecs for
GMCs). Gravitational focusing further increases the
cross-section. Also, the small scale height of GMCs implies
essentially two dimensional interactions in the plane of the disk,
increasing the collision rate relative to that for three
dimensions. We calculate collision timescales short enough to allow a
viable theory of collision induced star formation to be considered.

In summary, in this model, self-gravitating gas disks fragment into
bound gas clouds. This process is driven either by gravitational,
thermal or Parker instabilities, or the influence of stellar spiral
density waves on the gas. These bound clouds, either atomic or
molecular, are relatively long-lived, being supported by static and
turbulent magnetic pressure. The latter may be produced by
dynamically-regulated low mass star formation (McKee 1999). We
hypothesize a fraction of cloud collisions lead to compression of
localized regions of the clouds. These regions, if magnetically
supercritical, collapse rapidly to form stars, including high mass OB
stars. The bulk of Galactic disk stars are thought to form via this
``burst''-mode (Lada et al 1993). Thus, the rate limiting step for
star formation is not the formation of bound clouds, but the
compression of these, or parts of these, in cloud-cloud
collisions. Therefore at any particular time, most of the bound gas is
not actively undergoing star formation.

\begin{figure}
\plotone{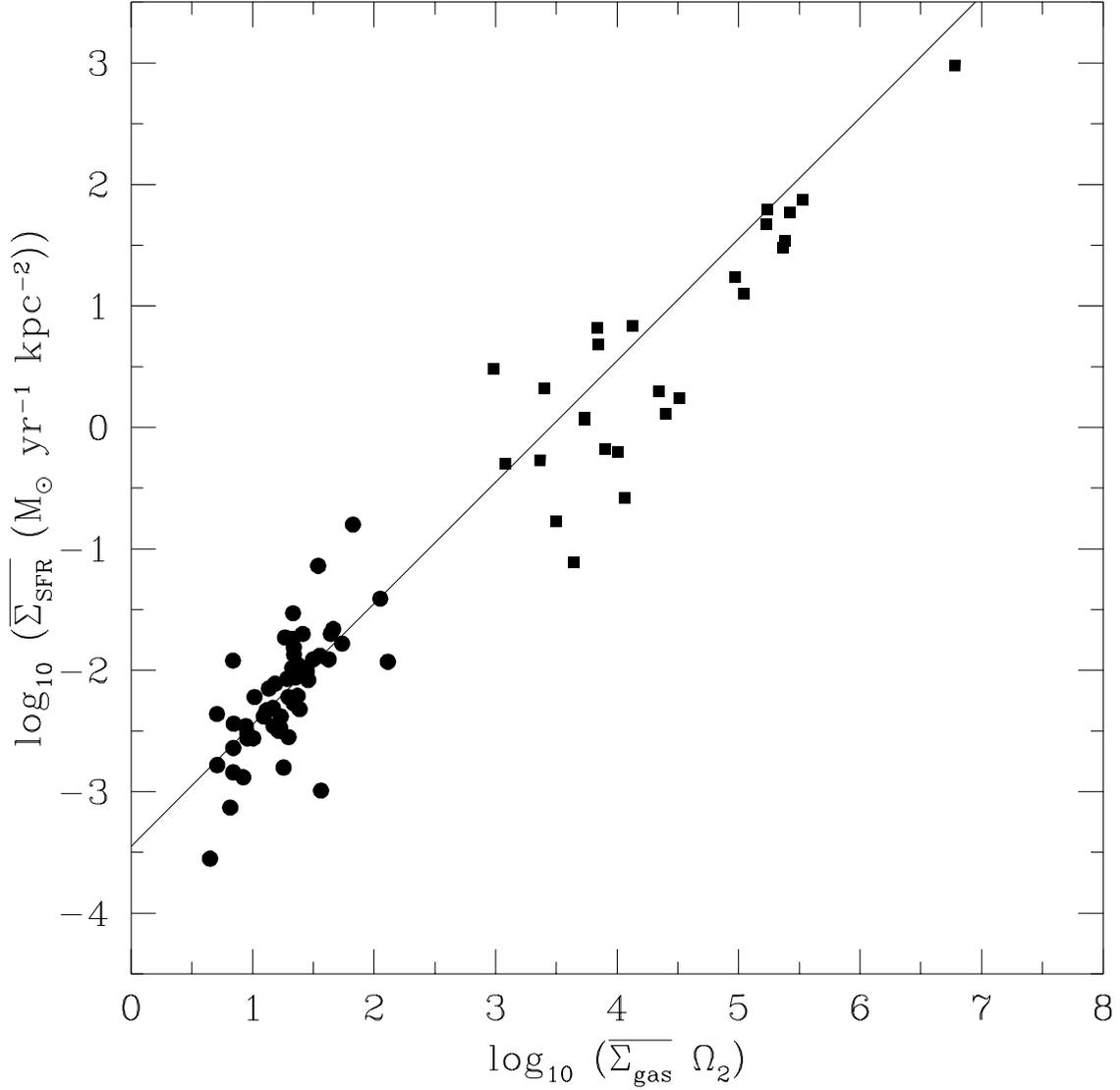}
\figcaption{
Modified Schmidt law : $\overline{\Sigma_{SFR}}\propto \overline{\Sigma_{gas}} \Omega_{2}$. From Kennicutt (1998). Data are disk averaged quantities for normal galactic disks ({\it solid circles}) and circumnuclear starburst disks ({\it solid squares}). The line is a median fit to the normal galactic disk sample, with the slope fixed at unity as predicted by equation (\ref{sfr7}). Systematic uncertainties between the normalization of the normal and starburst samples are of the order of a factor of two.
\label{fig:ken2}}
\end{figure}

Specifically, we have considered an idealized, single mass population
of gravitationally bound gas clouds, orbiting in an axisymmetric, thin
disk. Using the result of Gammie et al (1991) for the cloud velocity
dispersion, we predict radial gas distributions, dependent on the
Toomre $Q$ stability parameter (equation \ref{radialgas}). Applying
our principal hypothesis, that cloud collisions trigger the majority
of disk star formation, using the collision cross-section results of
Gammie et al (1991) and with the assumption star formation
self-regulates ($Q\sim {\cal O}(1)$), we predict enhanced cloud
collision rates and a SFR law of the form $\Sigma_{SFR}(R)\simeq1.5
\epsilon f_{sf} f_{G} Q^{-1} \Sigma_{gas} \Omega (1-0.7 \beta)$
(equation \ref{sfr3}). For flat rotation curves ($\beta=0$), this
result is in agreement with the disk averaged data of K98 (figure
\ref{fig:ken2}). Although uncertain, our estimates of the total SFR in
the Milky Way and for typical starburst systems are consistent with
observations. We predict a B-band Tully-Fisher relation of the form
$L_B \propto v_{circ}^{7/3}$, in agreement with observations (Burstein
et al 1995; Strauss \& Willick 1995).

\begin{figure}
\epsscale{0.75}
\plotone{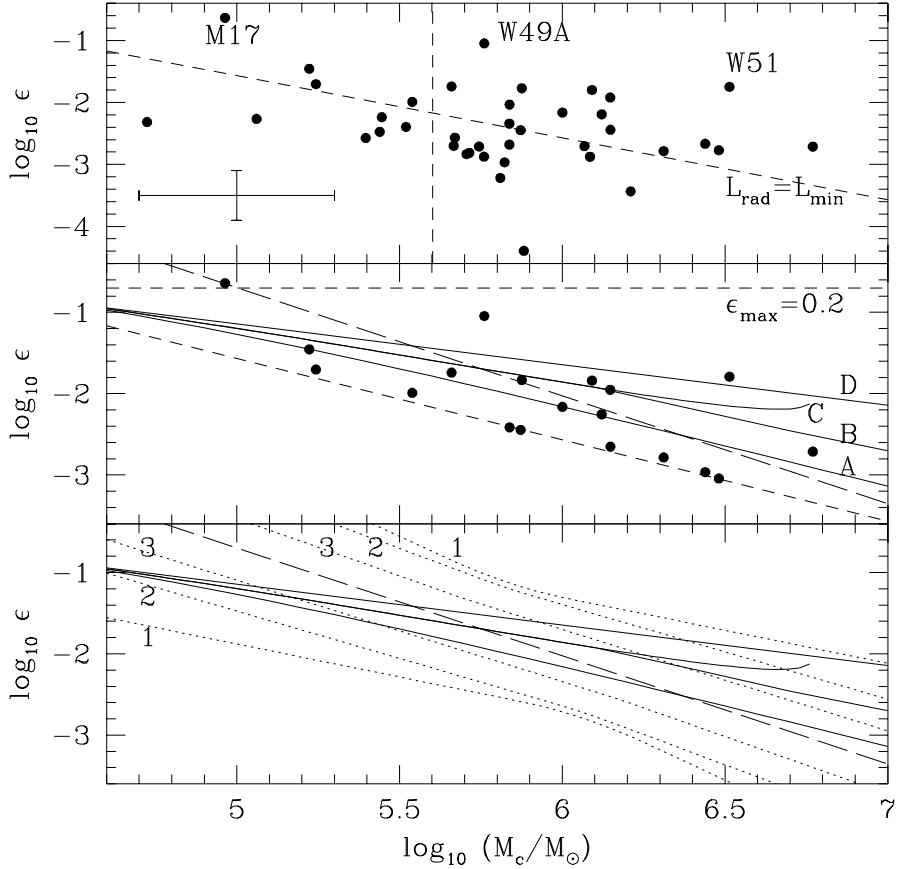}
\figcaption{ 
Star formation efficiency, $\epsilon$, versus cloud mass,
$M_c$. (a) {\it Top:} Full sample of 39 clouds associated with 83 HII
regions. The cross shows typical errors of 0.3 in ${\rm log_{10}}\: M_c$ and
0.4 in ${\rm log_{10}}\: \epsilon$. The {\it diagonal dashed} line shows the
efficiency a cloud would have if associated with an HII region of
luminosity $L_{min}=400\:{\rm Jy\:kpc^{2}}$. The data are incomplete
below this line. The {\it vertical dashed} line shows the molecular cloud
survey completeness boundary at $M_c=4\times 10^{5}\:{\rm M_{\odot}}$
(Williams \& McKee 1997). (b) {\it Middle:} Complete sample of 19
clouds associated with individual HII regions, each with
$L_{rad}>L_{min}$. The completeness boundary is again shown by the
{\it diagonal dashed} line. The adopted value of $\epsilon_{max}=0.2$ is
shown by the {\it horizontal dashed} line. The best linear fit to
this data in logarithmic space is shown by the {\it long dashed line}. {\it Solid} lines show various model predictions: A - Collision induced star
formation; B - Stochastic star formation (Williams \& McKee 1997); C -
Stochastic star formation with $\epsilon_{max}(M)={\rm constant}$; D -
Uniform distribution of $\epsilon$ in logarithmic space. (c) {\it
Bottom:} 95\% Confidence intervals on the best linear fit ({\it long dashed}
line) to the data in logarithmic space are shown by the {\it dotted} lines:
1 - limit for the existing 19 data points, with errors shown by cross
in (a); 2 - limit for hypothetical data set of 190 clouds with the
same distribution as the existing 19; 3 - limit for these 190 clouds
with typical errors half of those shown by cross in (a). Note,
although these limits are based on the (poor) assumption of a normal
distribution of the data about the best fit line, and are only limits
on linear fits (hence the deviation at low $M_c$), this figure still illustrates that with ten times more
data and with errors reduced by a factor of two, one can hope to
distinguish between the different models ({\it solid} lines, as in (b)).
\label{fig:eff}}
\end{figure}

This theory is to be further scrutinized to discriminate between it
and other star formation mechanisms. To this end we have proposed
several tests. We predict statistically enhanced SFRs in regions of
large negative circular velocity gradients, where the shear rate is
increased, and regions of increased cloud velocity
dispersion. Similarly, decrements are predicted in regions of large
positive circular velocity gradients, which reduce the amount of
shear. Future observations (e.g. Martin \& Kennicutt 2000) of SFR, gas
and circular velocity profiles of large samples of disk galaxies
should allow for statistically significant tests of our proposed SFR
law, and in particular the dependence on the circular velocity
gradients and cloud velocity dispersion. However, these tests will be
complicated by possible variations in the likelihood of collision
induced star formation with collision velocity. The results of
numerical simulations may be necessary to account for this effect. We
also predict star formation efficiency, $\epsilon$, linearly averaged,
decreases with increasing cloud mass as $\left<\epsilon\right>\propto
M_c^{-0.6}$. Figure \ref{fig:eff} shows model predictions for
$\epsilon$, logarithmically averaged over its distribution, and
comparison to observations. Larger and deeper surveys of HII regions
and GMCs, including their atomic components, are required to improve
the significance of this test.

Undoubtedly our model is an extremely simplified description of the
actual star formation process. We have presented an idealized theory
in which all star formation is triggered by cloud collisions, however
other processes, such as spontaneous star formation, self-triggering
and triggering by density waves undoubtedly operate at some level. For
the results of the collision induced theory to be valid, we require
that the majority of (high mass) star formation is initially triggered
by this process. The basic theory needs modification where there is a
tight correlation of star formation with large scale density waves,
allowing for the duration clouds spend in the density wave, and the
degree of spatial concentration.

The theory can be improved by numerical calculation of collision rates
in a many body system, rather than relying on simple two body
interaction rates. Numerical simulation of cloud collisions
(e.g. Klein \& Woods 1998) may provide insight into the details of how
a magnetically supercritical region can be produced from the collision
of two magnetically subcritical clouds. The parameter space for the
outcome of collisions with different initial conditions is also being
probed by simulation (Klein, private communication). These simulations
will constrain the probability, $f_{sf}$, for star formation to result
from typical cloud-cloud collisions.

This theory can be applied to analytic models (e.g. Shore \& Ferrini
1995; Silk 2000) and simulations (e.g. Curir \& Mazzei 1998; Weil, Eke
\& Efstathiou 1998) of disk galaxy formation and evolution, for
comparison to cosmological SFR data.

\acknowledgements We thank Chris McKee for many hours of stimulating
 discussion and much input. We also thank Leo Blitz, Andrew Cumming,
 Alex Filippenko, Rob Kennicutt, Richard Klein, Chris Matzner, Antonio
 Parravano, Evan Scannapieco, Joe Silk and an anonymous referee for
 helpful comments.

\newpage

\newpage

\begin{deluxetable}{cccc} 
\small
\tablecaption{Cloud properties\label{tab:prop}}
\tablewidth{0pt}
\tablehead{
\colhead{Property} & \colhead{Formula or Source} & \colhead{Galactic disk} & \colhead{Circumnuclear disk}\\
}
\startdata
$M_{c}$ & obs./magnetic critical mass & $\sim 5 \times10^{5}\:{\rm M_{\odot}}$ & $\sim 1 \times10^{8}\:{\rm M_{\odot}}$\tablenotemark{a}\\
$R$ & observation & $\sim 4000\:{\rm pc}$ & $\sim 600\:{\rm pc}$\tablenotemark{b}\\
$v_{circ}$ & observation & $225\:{\rm km/s}$ & $300\:{\rm km/s}$\\
$r_{t}$ & $=(2M_{c}/M_{gal})^{1/3}R$ & $\sim 100 \:{\rm pc}$ & $\sim 100 \:{\rm pc}$\\
$r_{c}$ & observation & $\sim20\:{\rm pc}$\tablenotemark{c} & $<100\:{\rm pc}$\tablenotemark{d}\\
$c_s$ & Alfven velocity& $\sim 1.5\:{\rm km/s}$ & uncertain\\
$v_{rel}$ & $\sim 1.6r_t \Omega + (2GM/r_c)^{1/2}$ & $13\:{\rm km/s}$ & $\sim 200\:{\rm km/s}$ \\
$\bar{n}_{H}$ & $0.75 M_c/(\frac{4}{3}\pi r_c^3)$& $\sim 450\:{\rm cm^{-3}}$ & $\sim 1.7\times10^{4}\:{\rm cm^{-3}}$\\
\enddata
\tablenotetext{a}{DS98 are only able to resolve a few of the largest bound clumps, of mass $\sim 10^{9}\:{\rm M_{\odot}}$, in the circumnuclear disks. By analogy with GMCs in normal disks, we take the typical mass to be an order of magnitude less than this maximum.}
\tablenotetext{b}{This is the mean value of $R_{1}$, the inner disk half intensity (of CO flux) radius, from the sample of DS98.}
\tablenotetext{c}{We take this fiducial value for consistency with the clouds modeled by Gammie et al (1991). Real clouds of this mass will probably be somewhat more extended, particularly allowing for HI envelopes.}
\tablenotetext{d}{In circumnuclear disks $r_c$ is uncertain. For the calculations which require a definite value, we take $r_c=35$ pc}
\end{deluxetable}

\newpage

\begin{deluxetable}{cccc} 
\small
\tablecaption{Cloud timescales\label{tab:times}}
\tablewidth{0pt}
\tablehead{
\colhead{Process} & \colhead{Formula or Reference} & \colhead{Time (years)} & \colhead{Time (years)}\\
\colhead{} & \colhead{} & \colhead{Galactic disk} & \colhead{Circumnuclear disk}
}
\startdata
Orbital Period, $t_{orb}$ & $2\pi R/v_{circ}$ & $\sim110\times10^{6}$ & $\sim12\times10^{6}$\\
Free-fall, $t_{ff}$ & $(3\pi/32G\rho_{gas})^{0.5}\simeq 4.33\times 10^{7} \bar{n}^{-1/2}_{H}$ & $2.0\times 10^6$ & $0.3\times 10^6$\\
Atomic to Molecular, $t_{conv}$ & $\sim(R_{mol}n_{HI})^{-1}$ & ${\rm few}\:\times 10^{6}$ & $\sim0.1\times 10^{6}$ \\
Ambipolar diffusion\tablenotemark{a}, $t_{AD}$ & $\simeq 15t_{ff}$ & $\gtrsim 30 \times 10^{6}$ & $\gtrsim 5 \times 10^{6}$\\
Collision\tablenotemark{b}, $t_{coll}$ & $\sim(3.2 \Omega {\cal N}_{A} r_{t}^{2}f_{G})^{-1}\sim t_{orb}/5$ & $\sim22 \times 10^{6}$ & $\sim2.4 \times 10^{6}$\\
Destruction, $t_{dest}$ & Williams \& McKee (1997) & $\sim30\times10^{6}$ & uncertain\\
Lifetime\tablenotemark{c}, $t_{exist}$ & $\gtrsim t_{dest}+{\rm min}(t_{AD},t_{coll})$ & $\gtrsim50 \times 10^{6}$ & $\gtrsim10 \times 10^{6}$\\
Alfven Crossing, $t_{cross}$ & $2r_c/c_s$ & $\simeq 25\times 10^{6}$ & uncertain\\
Impact time, $t_{imp}$ & $2r_c/v_{rel}$ & $\simeq 2.5\times 10^{6}$ & $\sim 0.3\times 10^{6}$\\
\enddata
\tablenotetext{a}{Note that the estimate of $t_{AD}$ is based on ionization solely from cosmic rays (see McKee 1999, Eq. 2 \& 89). The inhomogeneous nature of interstellar gas clouds means that UV radiation is much more penetrating than in the homogeneous case and that most of the gas mass of clouds is probably at a higher level of ionization and hence subject to longer ambipolar diffusion timescales than the above estimate.}
\tablenotetext{b}{This collision timescale is sensitive to the approximation of a cloud population with single cloud mass, $M_c$. The time between collisions which cause star formation is $f_{sf}^{-1} t_{coll}$.}
\tablenotetext{c}{This is the lifetime of a gravitationally bound cloud, not explicitly a molecular cloud. Upper limits of $\sim 10^{8}$ years (e.g. Blitz \& Williams 1999) are quoted for GMC lifetimes. However, bound clouds, ignoring the atomic/molecular distinction, may live much longer.}
\end{deluxetable}

\end{document}